\documentclass[aps,showkeys,prb,twoside,final,floatfix,%
twocolumn,showpacs]{revtex4}
\usepackage{amsmath,amsfonts}
\usepackage[dvips]{graphicx,color}

\usepackage{dcolumn}
\usepackage{txfonts}
\begin{document}
\title{
\begin{center}
{\normalsize PHYSICAL REVIEW B 73, 184424 (2006)}
\end{center}
Berezinski\v\i--Kosterlitz--Thouless transition
in two--dimensional lattice gas models}
\author{Hassan CHAMATI}
\altaffiliation[Present and permanent address:]{ Institute of Solid State Physics, 
72 Tzarigradsko Chauss\'ee, 1784 Sofia, Bulgaria.}
\affiliation{
Unit\'a CNISM e
Dipartimento di Fisica "A. Volta", Universit\`{a} di Pavia,
via A. Bassi 6, I-27100 Pavia, ITALY}
\author{Silvano ROMANO}
\altaffiliation[Corresponding author.]{Electronic address: Silvano.Romano@pv.infn.it}
\affiliation{
Unit\'a CNISM e
Dipartimento di Fisica "A. Volta", Universit\`{a} di Pavia,
via A. Bassi 6, I-27100 Pavia, ITALY}

\begin{abstract}
We have considered two classical lattice--gas models, consisting of
particles that carry multicomponent magnetic momenta, and associated
with a two--dimensional square lattices; each site can host one particle
at most, thus implicitly allowing for hard--core repulsion; the pair
interaction, restricted to nearest neighbors, is ferromagnetic and
involves  only two components.
The case of zero chemical potential
has been investigated by Grand--Canonical Monte Carlo simulations; the
fluctuating occupation numbers now give rise to additional fluid--like
observables in comparison with the usual saturated--lattice situation;
these were investigated and their possible influence on the critical
behaviour was discussed. Our results show that the present model
supports a Berezinski\v\i--Kosterlitz--Thouless phase transition with a
transition temperature lower than that of the saturated lattice
counterpart due to the presence of ``vacancies''; comparisons were also
made with similar models studied in the literature.
\end{abstract}


\maketitle

\section{Introduction}
Planar rotators are models with potential interactions involving
two--component spins; on the other hand, in the XY model, the spins have
three components, only two of which are involved in the interaction, and
the XY model can be regarded as an extremely anisotropic (easy--plane)
Heisenberg model.  Actually, the terminological convention adopted here
as well as by other Authors is not always followed, and the name ``XY
model'' is sometimes used in the Literature to indicate planar rotators;
on the other hand, both models are known to produce the same
universality class. The two named models have been extensively studied,
and possess  a rich variety of applications in Statistical as well as
Condensed Matter Physics \cite{rKT1,gulacsi1998,nelson2002}.

These models have been especially studied in their saturated-lattice
(SL) version, where each site is occupied by a particle; as for
symbols, classical SL spin models are first defined here: we
consider a classical system, consisting of $n-$component unit
vectors $\mathbf{u}_k$ (usually $n=2,3$), associated with a
$d-$dimensional (bipartite) lattice $\mathbb{Z}^d$ (mostly $d=2$);
let $\mathbf{x}_k$ denote dimensionless coordinates of the  lattice
sites, and let $u_{k,\alpha}$ denote Cartesian spin components with
respect to an orthonormal basis $\mathbf{e}_{\alpha}$; particle
orientations are parameterized by usual polar angles $\{\varphi_j\}$
($n=2$) or spherical ones $\{(\theta_j,\phi_j)\}$ ($n=3$);

The interaction potentials considered here are  assumed to be
translationally invariant, restricted to nearest--neighbours,
ferromagnetic (FM); they are, in general, anisotropic in spin space,
and possess $O(2)$ symmetry at least:
\begin{equation}\label{eq01a}
\Psi_{jk}=-\epsilon \Omega_{jk},
\end{equation}
where
\begin{eqnarray}\label{eq01b}
\Omega_{jk}&=&a u_{j,n} u_{k,n} + b \sum_{\alpha<n} u_{j,\alpha}
u_{k,\alpha}; \nonumber \\
& &\qquad \epsilon >0,~0 \leq a \leq b,~b >0, ~\max(a,b)=1.
\end{eqnarray}
Here $\epsilon$ is a positive quantity setting energy and
temperature scales (i.e. $T=k_B t/\epsilon$, where $t$ denotes the
absolute temperature); the corresponding (scaled) Hamiltonian reads
\begin{equation}
\Lambda = -\sum_{\{j<k\}}\Omega_{jk},
\label{eq01c}
\end{equation}
where $\sum_{\{j<k\}}$ is restricted to nearest neighbours, with
each distinct pair being counted once. When $d=1,2$, potential
models defined by Eq. (\ref{eq01b}) produce orientational disorder at
all finite temperatures (thus no orientational ordering transition
at finite temperature) \cite{rSin,rGeor}; in the following, the
discussion shall be essentially restricted to the two cases
$n=2,~a=b=1$ (PR) and $n=3,~a=0,~b=1$ (XY), respectively.

It is by now well known that the PR model defined by $d=2$ supports
a transition to a low--temperature phase with slow
(inverse--power--law) decay of the correlation function and infinite
susceptibility; this is the very extensively studied
Berezinski\v\i--Kosterlitz--Thouless (BKT) transition (see, e.g.
Refs. \cite{rKT1,gulacsi1998,nelson2002,rKT0,rKT2,rMH2005} and
others quoted therein), whose existence was proven rigorously by
Fr\"ohlich and Spencer \cite{rKT0}: the transition temperature 
was estimated to be $\Theta_{PR}=0.88 \pm 0.01$ \cite{rKT2}, 
and a more recent and more refined 
result is $\Theta_{PR}=0.8929$ \cite{rMH2005};  the
specific heat exhibits a maximum at a higher temperature
\cite{rKT2,tobochnik1979}. Moreover, anisotropic models defined by
$d=2,~n=3,~0 \le a <b$ have been studied as well and proven to
support a BKT transition when the ratio $a/b$ is sufficiently small
\cite{rfl0}; an estimate of the transition temperature for the XY
case is given by $\Theta_{XY}=0.700\pm 0.005$
\cite{rfl1,rfl1a,rfl2}.

Lattice--gas (LG) extensions of the SL models can be defined as
well, where each lattice site hosts one particle at most, and site
occupation is also controlled by the chemical potential $\mu$; such
models have often been used in connection with alloys and
absorption; this methodology somehow allows for pressure and density
effects. Lattice--gas (LG) extensions of the continuous--spin
potential model considered here are defined by Hamiltonians
\begin{equation}\label{eq02}
\Lambda = \sum_{\{j<k\}} \nu_j \nu_k (\lambda - \Omega_{jk})- \mu N
,~\qquad N=\sum_k \nu_k,
\end{equation}
where $\nu_k=0,1$ denotes occupation numbers; notice that $\lambda
\leq 0$ reinforces the orientation--dependent term, whereas $\lambda
>0$ opposes it, and that a finite value of $\lambda$ only becomes
immaterial in the SL limit $\mu \rightarrow + \infty$.

For a square lattice, the SL--PR model produces a low--temperature
BKT transition; the existence of such a transition for the LG
counterpart has been proven rigorously as well \cite{gruber2002}. In
addition to the named rigorous results, still comparatively little
is known about the above magnetic LG models with continuous spins,
in marked contrast to the vast amount of information available for
their SL counterparts; for example, as far as we could check in the
Literature, their simulation study only started at the end of the
1990's \cite{romano1999,romano2000}.

A few years 
before  the existence of a BKT transition was 
proven for the LG--PR model \cite{gruber2002}, the same model had been 
investigated
by Monte Carlo simulation in the Grand--Canonical ensemble, and in
the absence of a purely positional interaction, i.e. $\lambda=0$
\cite{romano1999}. Simulations were carried out for $\mu=0.1$ and
$\mu=-0.2$, and showed that the BKT transition survives, even for
mildly negative $\mu$, and that the transition temperature is an
increasing function of $\mu$, in broad qualitative agreement with
previous Renormalization Group (RG)  studies \cite{rhe01,rhe02,rhe03}.

The Hamiltonian (Eq. (\ref{eq02})) can be interpreted as describing
a two--component system consisting of interconverting ``real''
($\nu_k=1$) and ``ghost'', ``virtual'' or ideal--gas particles
($\nu_k=0$); both kinds of particles have the same kinetic energy,
$\mu$ denotes the excess chemical potential of ``real'' particles
over ``ideal'' ones, and the total number of particles equals the
number of available lattice sites (semi--Grand--Canonical
interpretation).
The semi--Grand--Canonical interpretation was also used in early
studies of the two--dimensional planar rotator, carried out by the
Migdal--Kadanoff RG techniques, and aiming at two--dimensional
mixtures (films) of $^{3}$He and $^{4}$He \cite{rhe01,rhe02,rhe03}, where
non--magnetic impurities correspond to $^{3}$He. In addition to
phase separations, the results in Refs. \cite{rhe01,rhe02,rhe03} show
that, in a r\'egime of low fraction of non--magnetic impurities, the
paramagnetic phase undergoes a BKT transition.

Notice also that the above Hamiltonian (Eq. (\ref{eq02})) describes
a situation of {\em annealed} dilution; on the other hand, models in
the presence of {\em quenched} dilution, and hence the effect of
disorder on the BKT transition, have been investigated using the PR
model
\cite{rque01,rque02,leonel2003,berche2003,surungan2005,wysin2005}
and very recently the XY model \cite{wysin2005}; it was found that a
sufficiently weak disorder does not destroy the transition, which
survives up to a concentration of vacancies close to the percolation
threshold.

In this paper, we present an extensive study of the ferromagnetic
LG--PR and LG--XY models, whose Hamiltonian can be explicitly written as
\begin{equation}\label{eqpot01}
\Lambda = \sum_{\{j<k\}} \nu_j \nu_k \left[\lambda
- (u_{j,1}u_{k,1} + u_{j,2} u_{k,2}) \right]-
\mu N, 
\end{equation}
in order to gain insights into their  critical behaviour.
The models are further simplified by choosing $\lambda=0$, i.e.
no pure positional interactions; Let us also remark that
two--component spins are involved in the PR case, whereas XY
involves three--component spins but only two of their components are
involved in the interaction: in this sense the two models entail
different anchorings with respect to the horizontal plane in spin
space; moreover, an even greater variety of anchorings can be
realized via the recently introduced generalized XY models
\cite{rgenxy01,rgenxy02}.

The rest of the paper is organized as follows: in Section
\ref{simulation} we discuss details our simulation procedure. 
Section \ref{results} is devoted to the discussion of  simulation
results:
we have found evidence pointing to a BKT
transition, and used the  relevant finite--size scaling theory to locate
it. Possible effects of the
chemical potential on the nature of the transition are discussed
in Section \ref{conclusion}, which summarizes our
results.

\section{Monte Carlo simulations}\label{simulation} A detailed
treatment of Grand--Canonical simulations can be found in or via Refs.
\cite{romano1999,chamati2005a,rsim3}; the method outlined here has
already been used in our previous studies of other LG models
\cite{romano2000,chamati2005a,chamati2005b}. To avoid surface effects
simulations were carried out on periodically repeated samples,
consisting of $V=L^2$ sites, $L=40,80,120,160$; calculations were
carried out in cascade, in order of increasing reduced temperature $T$.

The two basic MC steps used here were Canonical and
semi--Grand--Canonical attempts; in addition two other features were
implemented \cite{rmult0,rHR}: (i) when a lattice sites was visited,
Canonical or semi--Grand--Canonical steps were randomly chosen with
probabilities ${\cal P}_{\rm can}$ and ${\cal P}_{\rm GC}$,
respectively; we used ${\cal P}_{\rm can}/{\cal P}_{\rm GC}=n-1$,
since spin orientation is defined by $(n-1)$ angles, versus one
occupation number and (ii) sublattice sweeps (checkerboard
decomposition) \cite{rmult0,rHR}; thus each sweep (or cycle)
consisted of $2V$ attempts, first $V$ attempts where the lattice
sites was chosen randomly, then $V/2$ attempts on lattice sites of
odd parity, and finally $V/2$ attempts on lattice sites on even
parity. Equilibration runs took between $25 \,000$ and $200\,000$
cycles, and production runs took between
$250\,000$ and $1\,000\,000$;
macrostep averages for evaluating statistical errors were taken over
$1\,000$ cycles. Different random--number generators were used, as
discussed in Ref. \cite{rHR}.

Computed observables included mean Hamiltonian per site and its
temperature derivative (specific heat at constant $\mu$ and $V$),
density and its derivatives with respect to temperature and chemical
potential, defined by
\begin{equation}\label{eqobs01}
H^*=\frac1V\left< \Lambda \right>,
\end{equation}
\begin{equation}\label{eqobs02}
\rho = \frac1V \left< N  \right>,~
\end{equation}
and by the fluctuation formulae (see e.g. \cite{rsim3})
\begin{equation}\label{eqfl01}
\rho_{T}=\left.\frac{\partial \rho}{\partial T}\right|_{\mu,V}~
= \frac1{VT^2} \left[\left<N\Lambda\right>
-\left<N\right>\left<\Lambda\right>\right],
\end{equation}
\begin{equation}\label{eqfl02}
\rho_{\mu}=\left.\frac{\partial \rho}{\partial \mu}\right|_{T,V}~
= \frac1{VT} \left[\left<N^2\right>-\left<N\right>^2\right],
\end{equation}
\begin{equation}\label{eqfl03}
\frac{C_{\mu,V}}{k_B} = \left.\frac1{k_B}\frac1V
\frac{\partial\left< \Lambda \right>}{\partial T}\right|_{\mu,V}
=\frac1{VT^2}
\left[\left<\Lambda^2\right>-\left<\Lambda\right>^2\right].
\end{equation}
We also calculated mean in--plane magnetic moment per site and
in--plane susceptibility defined by
\begin{equation}
M =\frac1V \left< \sqrt{\mathbf{F} \cdot \mathbf{F}} \right>
\end{equation}
and
\begin{equation}\label{eqchi2}
\chi = \frac1{VT}
\left\langle\mathbf{F}\cdot\mathbf{F}\right\rangle,
\end{equation}
where
\begin{equation}\label{eqfl07}
\mathbf{F}=\sum_k \nu_k \left( u_{k,1} \mathbf{e}_1+u_{k,2}\mathbf{e}_2 \right),
\end{equation}
taking into account only the in--plane components of the vector
spin.

A square sample of $V$ sites contains $2 V$ distinct
nearest--neighbouring pairs of lattice sites; we worked out pair
occupation probabilities, i.e. the mean fractions $R_{JK}$ of pairs
being both empty ($R_{ee}=\left<(1-\nu_j)(1-\nu_k)\right>$), both
occupied ($R_{oo}=\left<\nu_j\nu_k\right>$), or consisting of an
empty and an occupied site
($R_{eo}=\left<(1-\nu_j)\nu_k+(1-\nu_k)\nu_j\right>$). It should be
noted that $R_{ee}+R_{oo}+R_{eo}=1$.

Short-- and long--range positional correlations, were compared by
means of the excess quantities
\begin{equation}
R^{\prime}_{oo}=\ln \left( \frac{R_{oo}}{\rho^2} \right),~
R^{\prime \prime}_{oo}=R_{oo}-\rho^2,
\label{eqexcpos}
\end{equation}
collectively denoted by $R^*_{oo}$ (notice that these two
definitions entail comparable numerical values).

Quantities such as $\rho$, $\rho_T$, $\rho_{\mu}$ and the above pair
correlations $R_{JK}$ or $R^*_{oo}$ can be defined as ``fluid--like'',
in the sense that they all go over the trivial constants in the SL
limit. Let us also remark that some of the above definitions (e.g.
$C_{\mu,V}$ and $\rho_T$) involve the total potential energy both in
the stochastic variable and in the probability measure (``explicit''
dependence), whereas some other definitions, e.g. $\rho_{\mu}$ or
the quantities $R_{JK}$, involve the total potential energy only in
the probability measure (``implicit'' dependence).

\section{Simulation results}\label{results}

\begin{figure}[h!]
\resizebox{\columnwidth}{!}{\includegraphics{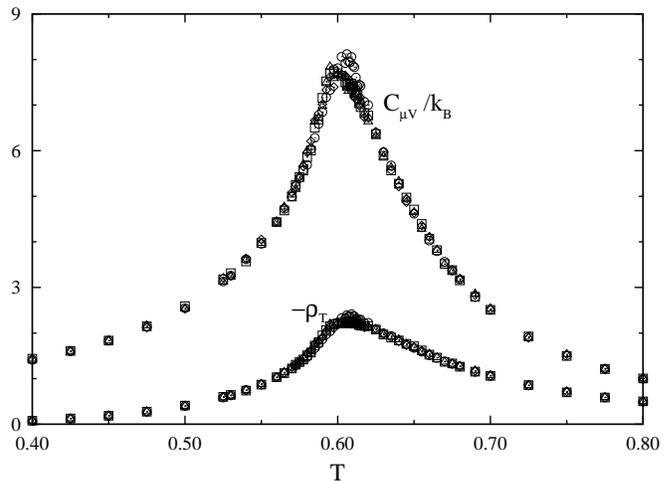}}
\caption{Simulation estimates for the specific heat per site and
for $-\rho_T$, obtained with different sample sizes: circles: $L=40$;
squares: $L=80$; triangles: $L=120$; diamonds: $160$. The value
$\mu=0$ was used in the present calculations.
Statistical errors for $C_{\mu,V}$, not reported, range between $1$ and
$3 \%$; otherwise, here and in the following Figures,
statistical errors fall within symbol sizes.} \label{cmv}
\end{figure}

We start by discussing the outcome of the
simulations performed for the LG--XY model. Results for a number of
observables, such as the mean energy per site $H^*$, and density
$\rho$ (not reported here) were found to evolve with temperature in
a smooth way, and to be independent of sample sizes;
the derivatives $C_{\mu,V}$ and $\rho_T$ (both plotted
on FIG. \ref{cmv}) showed a rather smooth trend with temperature,
and a ``sharp'' peak at $T \approx 0.6$, around which the
sample--size dependence of results became \textit{slightly} more
pronounced;
notice that neither the specific heat nor $\rho_T$
are exhibiting a divergence as a function of the temperature.

At all investigated temperatures simulation results for $M$ (not
reported here) exhibited the expected power--law decay with
increasing sample size, i.e. they were well fitted by the relation
\begin{equation} \label{eqaddres}
\ln M = -b_1 \ln L + b_0, \qquad ~b_1>0,
\end{equation}
where the ratio $b_1(T)/T$ was found to increase with temperature; a
spin--wave analysis of the SL--PR (Refs. \cite{rKT4,rKT4a} and those
quoted therein) predicts a similar sample--size dependence, in
agreement with simulation results for the present model;
furthermore, in the low--temperature limit $b_1$
becomes proportional to $T$.

\begin{figure}[h!]
\resizebox{\columnwidth}{!}{\includegraphics{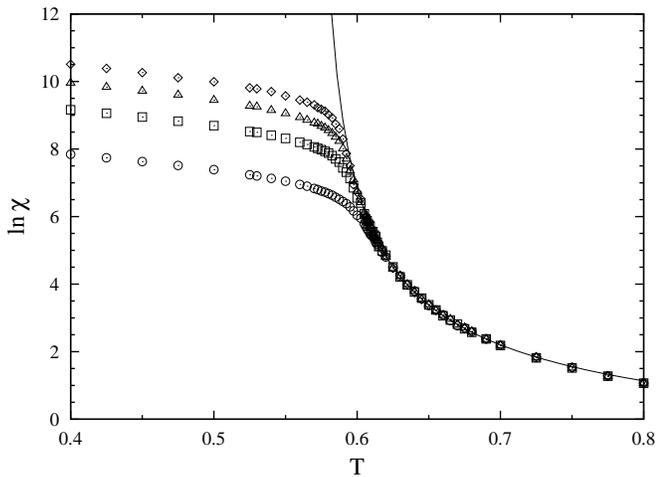}}
\caption{Simulation estimates for
the logarithm of the magnetic susceptibility $\chi$ versus temperature,
obtained with different sample sizes; same meaning of symbols as
in Fig. \ref{cmv}.
The solid curve is obtained by fitting to equation (\ref{chibkt}) with
data corresponding to $L=160$.}
\label{susceptibility}
\end{figure}

The behaviours of specific heat and susceptibility
(FIG. \ref{susceptibility}) 
suggest that the LG--XY model undergoes a BKT--like phase
transition. According to the BKT theory, in the thermodynamic
limit, the in--plane susceptibility, $\chi$, diverges exponentially
while approaching the transition temperature $\Theta$ in the high
temperature region \cite{gulacsi1998} i.e.
\begin{equation}\label{chibkt}
\chi\sim a_{\chi} \exp\left[b_{\chi}\left(T-\Theta\right)^{-\frac12}\right];
\qquad T\to\Theta^+
\end{equation}
and remains infinite in the low temperature region $T<\Theta$.

For a finite sample with volume $V=L^2$, the situation is different,
since the susceptibility obeys the constraint $\chi \le V/T$; thus
$\chi$ is always finite, and its exponential divergence in the
transition region is rounded, because of the important finite--size
effects. Actually, results for $\ln \chi$ versus temperature (FIG.
\ref{susceptibility}) were found to be independent of sample size
when $T \gtrsim 0.62$, and showed a recognizable increase with it (a
power--law dependence of $\chi$ on $L$) when $T \lesssim 0.6$.

In the critical region, i.e. $T\sim\Theta$, the correlation length
is of the same order as the linear size, $L$, of the system. In this
case, the exponential divergence disappears in the critical region,
but a reminiscence of the divergence can still be found in the
behaviour of $\chi$ at higher temperatures, where the correlation
length is still small compared to large linear system sizes and
finite--size effects can be neglected. We first fitted our MC
results obtained for the largest sample size and for temperatures in
the range $0.60 \lesssim T \lesssim 0.62$ to eq. (\ref{chibkt}), and
obtained evidence of the exponential divergence of the
susceptibility (see FIG. \ref{susceptibility}), and an estimated
transition temperature $\Theta=0.57\pm 0.01$, with
$a_{\chi}= 0.123 \pm 0.005$ and $b_{\chi}= 1.55 \pm 0.01$.

We checked our results by using data on a wider range of
temperatures extending up to $0.70$, and which yielded a consistent
result. In the following we refine our results using a more
elaborate method, namely the finite-size scaling theory.
Details on this procedure can be found in or via
reference \cite{rfl1a}.

In the temperature region where the singularity of $\chi$ is rounded
the finite--size scaling theory holds, i.e. the correlation length
$\xi$ is proportional to the size of the sample, $\xi\sim L$. From
the behaviour of the magnetic susceptibility $\chi\sim\xi^{2-\eta}$,
with $\eta=1/4$ \cite{gulacsi1998}, we end up with
\begin{equation}\label{chieta}
\chi(\Theta)\sim L^{2-\eta},
\end{equation}
where we have omitted the corrections arising from the presence of
the background (analytic) contributions to the finite--size scaling.

\begin{figure}[h!]
\resizebox{\columnwidth}{!}{\includegraphics{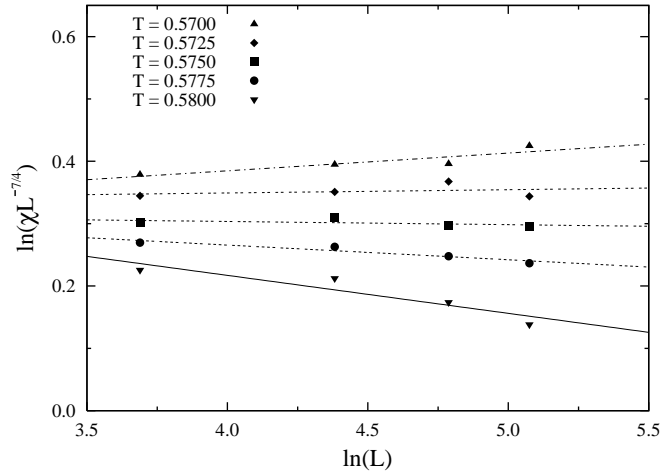}}
\caption{
Plot of $\ln \left(\chi L^{-7/4}\right)$ versus $\ln L$. Here the curve
closest to a horizontal straight line signals the transition.}
\label{eta}
\end{figure}

The temperature region around $T=0.57$, was analyzed in greater
detail, at first 
by carrying out a linear fit of $\ln \chi$ versus $\ln  L$
and extracting $\eta$ from the slope; the values
were found to be
 $\eta(T)=0.223\pm0.008$, $0.244\pm0.012$, $0.255\pm0.007$,
$0.273\pm0.005$, $0.308\pm0.015$ for $T=0.5700$, $0.5725$, $0.5750$,
$0.5775$, $0.5800$, respectively. A non--linear square fit, based on
equation (\ref{chieta}), was attempted as well, and yielded results
consistent with these ones; it also proved convenient for result 
visualization to plot 
$\ln (\chi L^{-7/4})$ versus $\ln L$
(FIG. \ref{eta}). 
Thus the transition temperature is estimated to be 
$\Theta=0.574\pm0.003$, in agreement with the above mentioned
result obtained by fitting the data of the susceptibility in the
high temperature region; 
at this temperature we expect the value of $\eta$ to be
$\frac14$ to within statistical errors.
%
%
This result shows that the
presence of ``vacancies'' in the sample reduce the transition
temperature by approximately $18\%$ compared to the SL transition
temperature, but does not change its nature.

The PR counterpart of the present model had been investigated in
Ref. \cite{romano1999}, where a maximum sample size corresponding to
$L=80$ was used. Moreover, in that paper the transition temperature
was estimated as the temperature where $\chi$ started exhibiting a
power--law dependency on $L$. Calculations had been carried out with
two different values of the chemical potential, $\mu=-0.2$ and
$\mu=+0.1$, respectively (see e.g. Table \ref{t01}); the transition
had been found to survive in both cases (even with a slightly
negative value), and the transition temperature had been found to
increase with increasing $\mu$, in qualitative agreement with
previous RG results \cite{romano1999}.

\begin{table}
\caption[]{
Comparison of the
transition temperatures, $\Theta_{PR}$, of the LG--PR model obtained in Ref.
\cite{romano1999} and those obtained using the finite--size scaling
presented in the present paper.}
\label{t01}
\begin{ruledtabular}
\begin{tabular}{rrr}
$\mu$ \ \   & $\Theta_{PR}$ (Ref. \cite{romano1999})
                              & $\Theta_{PR}$ (present work)  \\
\hline
$-0.2$       &  $0.73\pm0.01$   &      $0.71\pm0.01$     \\
 $0.1$       &  $0.79\pm0.01$   &      $0.75\pm0.01$     \\
\end{tabular}
\end{ruledtabular}
\end{table}

For further comparison we reexamined the results obtained in the
case of the LG--PR model in two ways. On the one hand, the above
finite--size analysis was applied to the simulation data produced in
Ref. \cite{romano1999}. The transition temperatures, reported in
Table \ref{t01}, corresponding to the two values of the chemical
potential used were found to be lower than the ones reported
previously. The behaviour of the $\mu$--dependence of the
temperature was similar for both methods. On the other, in order to
achieve a better comparison with the present case, simulations were
run {\em anew} for the case $\mu=0$, using same sample sizes as for
the LG--XY counterpart
($L=40, 80, 120, 160$). 
The thermodynamic
quantities of interest were found to behave qualitatively in a
similar fashion as those obtained in the framework of the XY model.

A detailed investigation of the susceptibility via the above
finite--size scaling procedure resulted in a BKT--like transition
with a transition temperature estimate of $0.733\pm0.003$,
corresponding to a particle density of $0.924\pm0.003$. The
transition temperature in this case was found to be $18\%$ lower
than the case of the planar rotator SL counterpart.

\begin{figure}[h!]
\resizebox{\columnwidth}{!}{\includegraphics{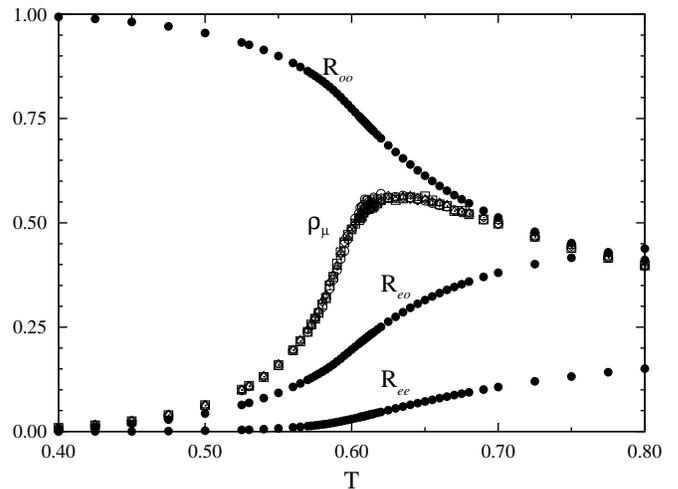}}
\caption[]{Simulation estimates for the three pair occupation
probabilities $R_{JK}$ obtained for a
two--dimensional sample with linear size $L=160$, 
along with simulation estimates for $\rho_\mu$, obtained with
different sample sizes; same meaning of symbols as in FIG.
\ref{cmv}.} \label{proba}
\end{figure}

\begin{figure}[h!]
\resizebox{\columnwidth}{!}{\includegraphics{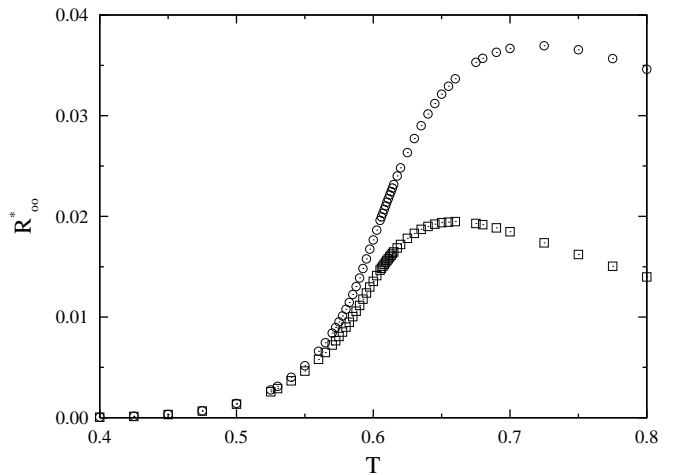}}
\caption{Simulation estimates for
the quantities $R^*_{oo}$; discrete symbols have been
used for simulation results obtained with $L=160$, and have
the following meanings: circles: $R^{\prime}_{oo}$; squares:
$R^{ \prime \prime}_{oo}$.}
\label{rstar}
\end{figure}

As for other fluid--like quantities obtained within the framework of LG--XY,
the plot of the derivative $\rho_\mu$
(FIG. \ref{proba}) against the temperature exhibited a smooth
evolution, a broad maximum at $T \approx 0.625$ and a comparatively
weak sample--size dependence;
pair occupation probabilities $R_{JK}$ were found to be essentially
independent of sample size, similarly to quantities such as $H^*$
and $\rho$; their results for the largest sample size $L=160$ are
presented in FIG.~\ref{proba} as well; the three quantities evolve with
temperature in a gradual and monotonic way, 
and their plots suggest inflection points between $T=0.6$ and $T=0.625$,
roughly corresponding
to the maximum in $\rho_{\mu}$. Short-- and long--range positional
correlations have been compared via the excess quantities
$R^*_{oo}$, whose simulation results for the largest sample size are
reported in FIG.~\ref{rstar}, showing a broad maximum well above the
transition temperature; notice also that $R^{*}_{oo}$ remains rather
small, reflecting the absence of a purely positional term in the
interaction potential. 
In general it has been found that the
obtained properties exhibit a recognizable similarity with with the
counterparts for the isotropic LG--PR model \cite{romano1999}. Some
other remarks are appropriate, concerning the temperature dependence
of fluid--like properties; let us recall that our potential model
contains no explicit positional interaction, i.e $\lambda=0$ in Eq.
(\ref{eqpot01}), so that positional properties are essentially
driven by the orientation-dependent pair potential. Similarities and
differences in the above plots for $C_{\mu,V}$ and $\rho_T$ 
on the one hand
(FIG.~\ref{cmv}), and $\rho_{\mu}$ or $R_{JK}$ (FIG.~\ref{proba})
on the other hand,
can then be correlated with the ''implicit'' or ``explicit''
dependencies of the named quantities on the potential energy (see
remarks at the end of the previous section);
the pair interaction energy changes most rapidly at $T\approx 0.6$,
and this is reflected by narrow peaks in $C_{\mu,V}$ and $\rho_T$;
conversely, quantities with ``implicit dependence'' on the
pair interaction energy are less affected by the change, as shown by
a broader peak at higher temperature $T\approx 0.625$.

\section{Concluding remarks}\label{conclusion}
In this paper, we have investigated two magnetic lattice--gas
models living on two--dimensional lattices, i.e.
PR and XY, respectively.
We have studied both lattice--gas models, in the absence of a pure
positional interaction, using semi--Grand--Canonical
Monte Carlo Simulation. In the case of zero chemical potential a number
of thermodynamic quantities observable, including some characteristic
of fluid systems, were computed. In general it has been found that the
obtained properties for both models exhibit a recognizable qualitative
similarity.

Results for the susceptibility yield consistent evidence of the
existence of a BKT transition, suggesting the estimates
$\Theta_{XY}=0.574\pm0.003$ and $\Theta_{PR}=0.733\pm0.003$ for the
transition temperatures, with corresponding densities at transition
$\rho_{XY} =0.918\pm0.004$ and $\rho_{PR}=0.924\pm0.003$, respectively. These
results are consistent with those reported in Ref. \cite{wysin2005}
obtained in the framework of models describing quenched dilution;
the ratio of the transition temperatures of the LG models to the SL ones
is approximately $0.82$. The peak of the specific
heat falls closer to $\Theta_{BKT}$ than for 
the SL counterpart; this can be interpreted as reflecting the fact that,
 since $\lambda=0$, the strengthening of short--range orientational 
correlations also brings about
a more rapid increase of density, and hence, in turn, of $H^*$.

As for the possible $\mu$--dependence of the present results, let us
first notice that, since $(\partial \rho /\partial \mu)_{T,V}>0$, it
is reasonable to expect that $\chi$, and hence $\Theta_{LG}$, are
increasing functions of $\mu$. On the other hand, the ground--state
energy for the SL counterpart is $-V_0$ per site, where $V_0=2$;
thus the lattice is essentially saturated when $T \leq 1$ and $\mu$
exceeds a few multiples of $V_0$ (say $\mu$ ranging between  $5$ and
$10V_0$); at the other extreme, when $\mu$ is negative and of
comparable magnitude, one expects an essentially empty lattice,
where no BKTLT should survive. 
Moreover, a BKTLT may still survive
if $\mu$ is negative but not too large in magnitude; this conjecture
is supported by RG studies for the LG--PR 
\cite{rhe01,rhe02,rhe03}, 
and was confirmed 
by the simulation results obtained in Ref. 
\cite{romano1999}; it also 
agrees with the conclusions of references
\cite{leonel2003,berche2003,surungan2005,wysin2005} for
two--dimensional magnets with quenched dilution;
for large negative $\mu$, the named RG treatments predict
phase separation,
i.e. a first--order transition between a BKT and a paramagnetic
phase, and eventually the disappearance of the BKT phase.  
We expect to
investigate the $\mu$--dependence of the transition temperature in a
more detailed way in future work.

\section*{Acknowledgements}
The present extensive calculations were carried out, on, among other
machines, workstations,  belonging to the Sezione di Pavia of
Istituto Nazionale di Fisica Nucleare (INFN); allocations of
computer time by the Computer Centre of Pavia University and CILEA
(Consorzio Interuniversitario Lombardo per l' Elaborazione
Automatica, Segrate -- Milan), as well as by CINECA (Centro
Interuniversitario Nord--Est di Calcolo Automatico, Casalecchio di
Reno - Bologna), are gratefully acknowledged. H. Chamati's stay at
Pavia University was made possible by a NATO--CNR fellowship;
financial support as well as scientific hospitality are gratefully
acknowledged; the authors also thank Prof. V. A. Zagrebnov
(CPT--CNRS and Universit\'e de la M\'editerran\'ee, Luminy,
Marseille, France) for helpful discussions.

\end{document}